# Film flip and transfer process to enhance light harvesting in ultrathin absorber films on specular back-reflectors

Asaf Kay[+,1], Barbara Scherrer[*,+,1], Yifat Piekner[2], Kirtiman Deo Malviya[1], Daniel A Grave[1], Hen Dotan[1] and Avner Rothschild[1]

[+]equal contribution

[1] Department of Materials Science and Engineering, Technion – Israel Institute of Technology, Haifa 32000, Israel

[2] The Nancy & Stephen Grand Technion Energy Program (GTEP), Technion – Israel Institute of Technology, Haifa 32000, Israel

[*]corresponding author: barbaras.emd.cloud@csu.iucc.ac.il

**Graphical abstract**

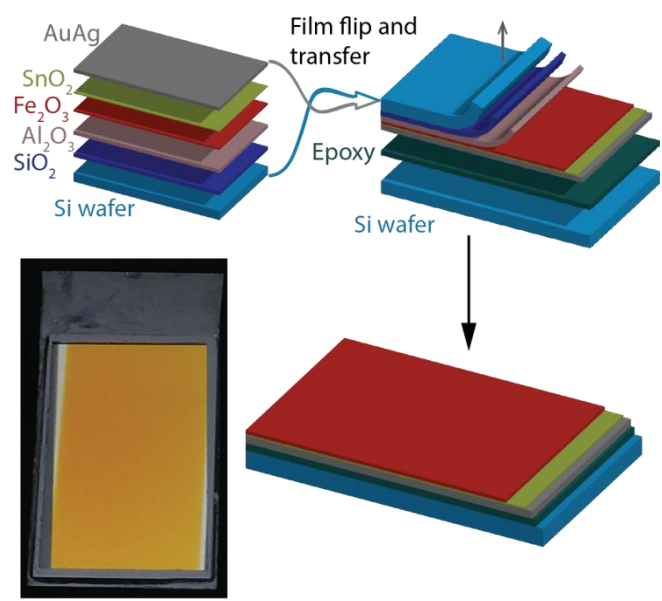

**Table of Contents**

We introduce a film flip and transfer process to enable fabrication of nanophotonic structures that leverage strong optical interference in ultrathin absorbers on metallic specular reflectors. This method allows for high temperature processing of metal-oxide absorbers without degradation of the metallic back reflector. We showcase this method by fabricating highly absorbing hematite ultrathin films on specular silver-gold back-reflectors for use in solar water splitting. The film flip and transfer process opens up a new route to attach thin film stacks onto a wide range of substrates including flexible or temperature sensitive materials.

**Abstract**

Optical interference is used to enhance light-matter interaction and harvest broadband light in ultrathin semiconductor absorber films on specular back-reflectors. However, the high-temperature processing in oxygen atmosphere required for oxide absorbers often degrades metallic back-reflectors and their specular reflectance. In order to overcome this problem, we present a newly developed film flip and transfer process that allows for high-temperature processing without degradation of the metallic back-reflector and without the need of passivation interlayers. The film flip and transfer process improves the performance of photoanodes for photoelectrochemical water splitting comprising ultrathin (< 20 nm) hematite ($\alpha$-Fe$_2$O$_3$) films on silver-gold alloy (90 at% Ag-10 at% Au) back-reflectors. We obtain specular back-reflectors with high reflectance below hematite films, which is necessary for maximizing the productive light absorption in the hematite film and minimizing non-productive absorption in the back-reflector. Furthermore, the film flip and transfer process opens up a new route to attach thin film stacks onto a wide range of substrates including flexible or temperature sensitive materials.

**Introduction**

Strong interference in ultrathin absorber layers on specular back-reflectors present a new paradigm in nanophotonics.[1–6] Significant absorption of broadband light can be achieved in layers only a few nanometers thick, with low sensitivity to the angle of incidence. However, metallic reflectors often degrade at high temperature and corrode in oxygen containing atmosphere. This is particularly problematic for silver, one of the best reflectors for the visible spectrum.[7] High reflectivity is necessary to minimize non-productive (wasted) absorption in the back-reflector. Detailed calculations reported elsewhere[6] identify silver as the best candidate for metallic back-reflectors designed to enhance light harvesting in the visible range. However, silver is known to tarnish upon air exposure and this accelerates at high temperatures as a result of thermal etching.[8] The silver surface oxidizes and the oxide evaporates, leading to surface roughening that damages the specular reflectance. This can lead to significant optical losses and substantial decrease in the amount of light absorbed in the ultrathin absorber (productive absorption). The deleterious tarnishing can be partially supressed by alloying silver with gold, as reported elsewhere.[6] Silver-gold alloys are more chemically stable than pristine silver, but their reflectivity degrades significantly at high gold concentrations ($\geq$ 15 at%), see **Figure S1**. A silver-gold alloy with 90 at% silver and 10 at% gold was found to present an optimal balance between stability and reflectivity.[6] Tarnishing was a key challenge in our previous work,[6] where strong interference in ultrathin hematite ($\alpha$-$Fe_2O_3$) films on silver-gold alloy (90 at% silver and 10 at% gold) back-reflectors was used for highly efficient photoanodes for photoelectrochemical water splitting. The silver-gold alloy layer was utilized both as a back-reflector as well as a back metallic contact. Formation of crystalline hematite with good photoelectrochemical properties require high temperature and oxygen atmosphere.[6,9–15] In order to passivate the silver-gold alloy layer and supress tarnishing during the subsequent hematite film deposition, three thin layers of $SnO_2$ deposited at increasing temperatures and oxygen pressures were used.[6] In order to leverage the strong interference effect, the thickness of the entire oxide stack must be tuned to the first resonance mode, for a silver-gold back-reflector this is ~20 nm depending on the optical constants of the films. This leaves very little room for the passivation interlayers. These layers also result in optical losses by absorbing light instead of the hematite absorber, giving rise to non-productive light absorption. Even with the use of passivation interlayers, the silver-gold layer lost some of its specular character due to surface roughening during the hematite deposition as shown in Figure S2 of reference.[6]

These challenges motivated us to explore an alternative process that avoids this problem by depositing the absorber layer before the silver-gold layer, and then flipping the structure upside down and removing the carrier substrate to expose the hematite layer. The idea is adapted from microfabrication processing where different techniques have been used to transfer epitaxial films and multilayer stacks from one substrate (the carrier substrate) to another.[16–25] In these studies, a range of methods were used to release thin films and transfer them including heat treatment, chemical etching, laser lift-off and exfoliation in dry or wet atmosphere. In our case, a stack consisting of hematite (absorber), $SnO_2$ (underlayer) and silver-gold (back-reflector) layers must be released from the carrier substrate, flipped upside down and transferred to another substrate, followed by removing the carrier substrate without breaking and damaging the functional layers. The combination of physically and chemically different materials (oxides and metal) that must be released, transferred and eventually exposed upside down complicates the entire process. In addition, no defects such as micro-cracks can be tolerated as these would expose the silver-gold layer to the harsh environment (alkaline solution) in which hematite photoanodes operate, resulting in corrosion that would eventually lead to device failure. These challenges render the reported film transfer methods inadequate to our purpose.

This work presents a new film flip and transfer process, which circumvents material incompatibility problems of ultrathin oxide absorbers on metallic back-reflectors, and is tailored to optimize the strong interference effect of such structures. We demonstrate this process with an ultrathin hematite film on silver-gold back-reflector serving as a photoanode for photoelectrochemical water splitting.

## Results and Discussion

### Flip film and transfer process

We reverse the deposition sequence starting with the hematite layer and ending with the silver-gold layer to avoid tarnishing that occurs due to thermal etching of the silver-gold alloy at high temperature in oxygen atmosphere. The deposition sequence on the carrier substrate is illustrated in **Figure 1** and in more detail in **Figure S2**. Prior to the deposition of the functional layers (hematite, $SnO_2$ and silver-gold alloy), a 250 nm thick thermal $SiO_2$ layer is grown on the Si wafer. The oxidized Si wafer is coated with two layers of 200 nm thick $Al_2O_3$, the first magnetron sputtered and second deposited by PLD (see **Figure 1**a, **Figure S2** and experimental section for details). The $SiO_2$ and $Al_2O_3$ layers serve as etch-stops in the subsequent etching processes that remove the carrier substrate and these layers to expose the hematite layer. Next, the hematite layer is deposited on the $Al_2O_3$ layer, followed by deposition of the 2 nm thick $SnO_2$ underlayer. Both the hematite and $SnO_2$ layers are deposited by pulsed laser deposition (PLD) at 300 °C in oxygen atmosphere. We explore different doping profiles in the hematite layers, including homogeneously doped hematite with 1 cat% Ti,[15] and heterogeneous doping profiles comprising 1 cat% Zn-doped hematite layer followed by undoped hematite layer followed by 1 cat% Ti-doped hematite layer.[13] The latter is called Zn-u-Ti in short and was shown to improve both the plateau photocurrent and the onset potential.[13,26] Following the deposition of the hematite and $SnO_2$ layers the entire stack is annealed in air at 500 °C for 2 h. Subsequently, the back-reflector layer is deposited from a silver-gold alloy (90 at% Ag with 10 at% Au). The silver-gold alloy was found in our previous study[6] to be more stable against oxidation than pristine silver while having high reflectance similar to pristine silver. Higher gold concentrations (above 10 at%) were found to degrade the reflectivity of the silver-gold alloy[6] see **Figure S1**. The 400 nm thick silver-gold alloy layer is deposited by sputtering in inert atmosphere (Ar) without heating.

Next, the entire stack on the carrier substrate is flipped and glued upside down onto another substrate (Si wafer), as illustrated in **Figure 1**b. Then, the carrier substrate, $SiO_2$ and $Al_2O_3$ layers are removed (**Figure 1**c), to expose the hematite layer (**Figure 1**d). The carrier substrate (Si wafer) and $SiO_2$ thermal oxide layer are removed by deep reactive ion etching (DRIE) and reactive ion etching (RIE), respectively. The $Al_2O_3$ layer is removed by wet etching in alkaline solution (1 M NaOH in deionized water). More details on the etching processes can be found in the experimental section. After removing the carrier substrate and the $SiO_2$ and $Al_2O_3$ layers, the hematite layer is exposed, as illustrated in **Figure 1**d. The film flip and transfer process results in specular hematite photoanodes, as shown in the photographs presented in **Figure 1**e for photoanodes with different doping profiles and hematite layer thicknesses. **Figure 1**f presents an exemplary cross-section TEM image of one specimen after the annealing step that precedes the silver-gold alloy deposition step. The hematite and $SnO_2$ layers are fully crystalline, and the interfaces between the different layers are sharp. Further microstructural characterizations are reported in the SI, showing that the photoanodes have smooth surfaces with only very few pinholes and no micro-cracks (see **Figure S3**). We also show cross-section TEM images taken of the entire stack (**Figure S4**), as well as defective specimens obtained with non-optimal process conditions (**Figure S5**). The film flip and transfer processes allowed the fabrication of ≥ 2 cm² photoactive area with no defects introduced by the process.

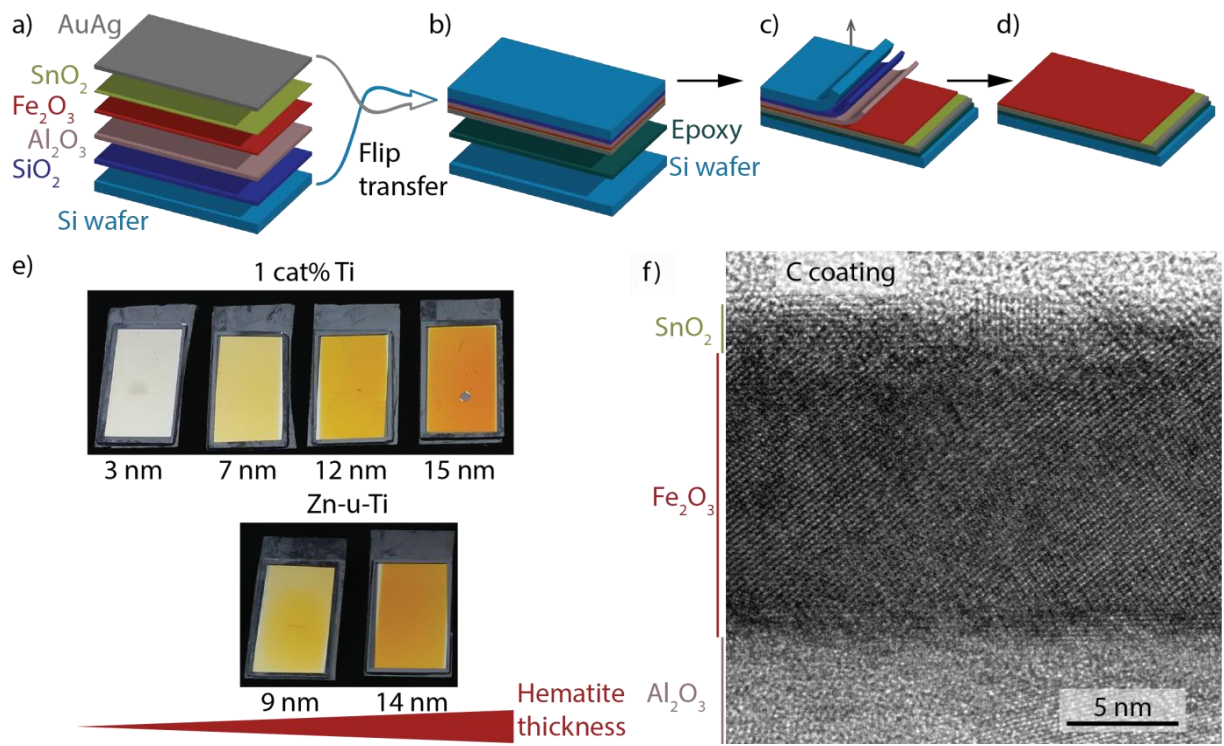

**Figure 1. Film flip and transfer process.** Schematic illustration of the film flip and transfer process, (a) starting from the deposition of different layers on the carrier substrate, (b) followed by flipping the deposited stack and gluing it upside down onto another substrate, (c) followed by removing the carrier substrate and $SiO_2$ and $Al_2O_3$ layers by dry and wet etching (d) to expose the hematite layer. (e) Photographs of photoanodes with different hematite layers. The doping profile and hematite thickness are depicted above and below the photographs, respectively. (f) Cross-sectional bright field TEM image of one of the photoanodes comprising a 14 nm-thick heterogeneously doped hematite layer with 2 nm $SnO_2$ underlayer after the annealing step and before the deposition of the silver-gold back-reflector.

Specular reflectance

Specular reflectance spectra of bare back-reflectors comprising of silver-gold alloy layers with 90 at% silver and 10 at% gold produced in different methods are shown in **Error! Reference source not found.**a. In the as-deposited state (grey curve) the bare back-reflector reflects more than 90% of the incident photons with wavelength above 400 nm, and the reflectance drops below 50% only at wavelengths below ~350 nm. The traditional bottom up deposition sequence (black curve), wherein the deposition of the silver-gold layer is followed by subsequent depositions of $SnO_2$ passivation interlayers and hematite layer at high temperature in oxygen atmosphere, results in visible tarnishing of the back-reflector due to thermal etching of the silver-gold alloy.[6,8] This leads to a considerable loss in reflectance across the entire spectrum (300-800 nm), see black curve in **Error! Reference source not found.**a. The film flip and transfer process presented here yields back-reflectors with high specular reflectance, similar to that of the silver-gold alloy in the as-deposited state (compare the green and grey curve in **Error! Reference source not found.**a). Specular reflectance spectra of photoanodes with different hematite layer thicknesses and doping profiles are shown in **Figure S6**. The total absorptance spectrum within the entire photoanode stack is calculated as $A(\lambda) = 1 - R(\lambda)$, where $R(\lambda)$ is the reflectance spectra. The film flip and transfer process results in highly specular hematite photoanodes, as shown in their photographs presented in **Figure 1**e.

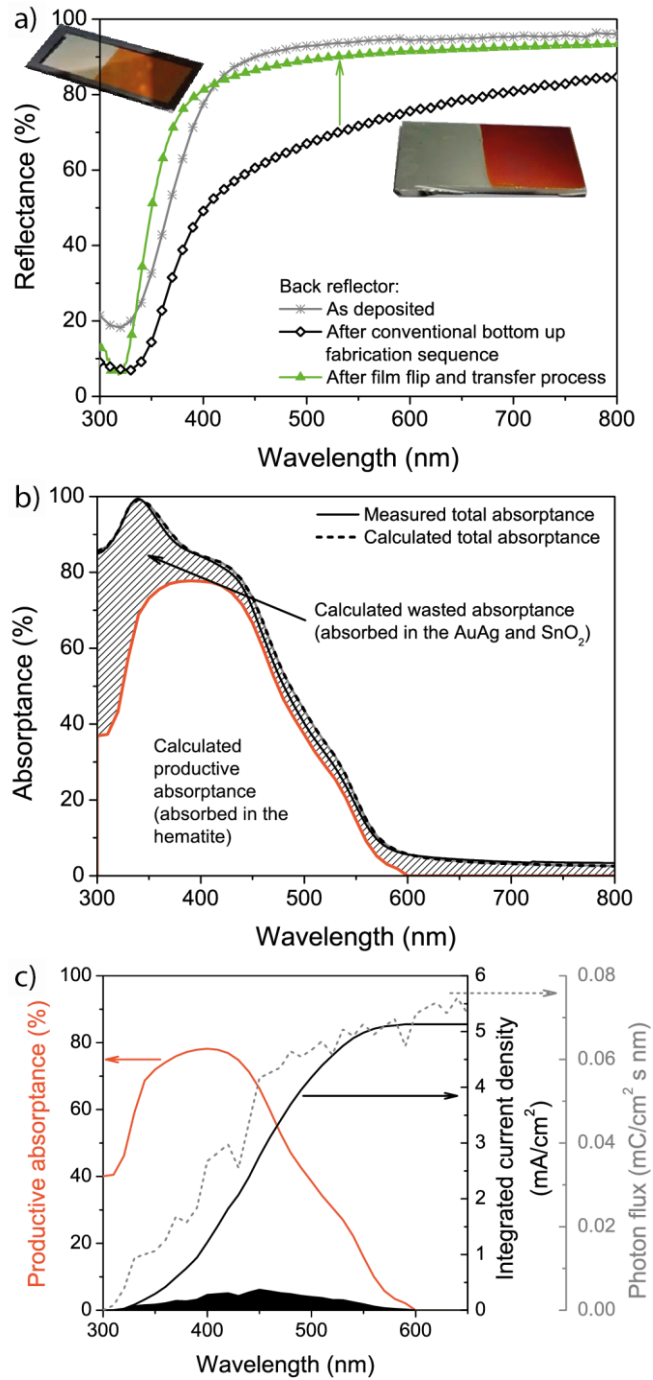

**Figure 2. Optical properties.** (a) Reflectance spectra of the silver-gold back-reflector in the as-deposited state (grey), after deposition of $SnO_2$ and hematite layers and annealing at 500 °C in the conventional photoanode fabrication sequence (black) and after the film flip and transfer process (green). (b) Total (measured and calculated) and productive (calculated) absorptance spectra (in air) of a photoanode comprising 8 nm thick 1 cat% Ti-doped hematite layer with a 2 nm thick $SnO_2$ underlayer on a silver-gold alloy (90 at% Ag and 10 at% Au) back-reflector, fabricated by the film flip and transfer process. (c) Calculated absorbed photon flux ($\phi_{abs}$) within the hematite layer of the photoanode featured in (b), with the photoanode immersed in water. The photon flux spectrum from the AAA solar simulator at the photoanode location is shown in the dashed grey curve. The shaded region shows the product of productive

absorptance (orange, in water) multiplied by the photon flux at the photoanode location, from which the integrated absorbed current density was calculated (black curve).

Productive light absorption

The total absorptance comprises of productive and non-productive contributions of absorption within the hematite layer or in other layers in the stack, respectively. The productive contribution, $A_{pro}(\lambda)$, is calculated with an optical simulator based on the transfer matrix method that accounts for the strong interference effect (**Figure S6**, **Figure S7**, **Figure S8** and **Figure S9**)[6] The optical constants of the different layers in the photoanode stack (hematite, $SnO_2$ underlayer and silver-gold back-reflector) were obtained by spectroscopic ellipsometry measurements (see **Figure S7**). The non-productive (wasted) contribution, $A_{wasted}(\lambda)$, is obtained by subtracting the productive contribution from the total absorptance, $A_{wasted}(\lambda) = A(\lambda) - A_{pro}(\lambda)$. **Error! Reference source not found.**b presents total and productive absorptance spectra of one of the photoanodes. The non-productive absorptance is marked by the patterned black background.

Using the calculated productive absorptance spectra of all the photoanodes, we calculate the absorbed photon flux within the hematite layer, $\phi_{abs} = \int \phi_{in}(\lambda) A_{pro}(\lambda) d\lambda$, where $\phi_{in}(\lambda)$ is the incident photon flux spectrum of the solar simulator (presented in **Figure S10**). **Error! Reference source not found.**c shows how $\phi_{abs}$ is calculated for the photoanode whose absorptance spectra are presented in **Error! Reference source not found.**b. The resultant $\phi_{abs}$ values are also expressed in terms of current density, so called the absorbed current density, $J_{abs} = q\phi_{abs}$, where $q$ is the electron charge. **Table 1** and **Figure S11** present the calculated $J_{abs}$ values for all the photoanodes in this study. The calculations presented in **Error! Reference source not found.** assume direct solar radiation at a normal angle of incidence in water. The effect of different angles of incidence is presented in **Figure S12**, showing little dependence in the range of 0° (normal incidence) to 50°. The broadband response and low sensitivity to the angle of incidence are important attributes of the resonant light trapping method that make it attractive for sunlight harvesting.[1,6]

The highest productive light absorption was achieved for the photoanode with a hematite layer thickness of 20 nm, reaching $J_{abs}$ = 9.5 mA/cm² for solar simulated illumination. This corresponds to 73 % of the theoretical limit for a thick hematite layer without reflection, which is quite remarkable considering the small thickness of this layer (20 nm). For comparison, the same hematite layer on transparent substrate would reach $J_{abs}$ of only 3.4 mA/cm$^2$, indicating a gain of a factor of 2.8 in productive light absorption for this film thickness.

Photoelectrochemical performance

In order to demonstrate the effect of our film flip and transfer process on the photoelectrochemical performance, linear sweep voltammetry (LSV) measurements of the ultrathin film hematite photoanodes were carried out in the dark and under front solar simulated illumination. The LSV measurements were done in alkaline solution (1 M NaOH in deionized water) with no sacrificial reagents. The results are presented in **Figure S13**, and the respective photocurrent (light current minus dark current) voltammograms are depicted in **Figure 3**. The dark current for all specimens is very low (**Figure S13**), showing the stability of the silver-gold back-reflector. Of all the homogenously Ti-doped (1 cat%) hematite photoanodes, with hematite thickness ranging from 3 to 15 nm, the one with a hematite thickness of 8 nm shows the highest photocurrent, reaching 0.77 mA/cm$^2$ at the reversible potential of water oxidation (1.23 V vs. RHE). The onset potential for the bare Ti-doped hematite photoanodes is slightly above 1.0 V vs. RHE (**Figure 3**a). With the addition of an ultrathin (~2 nm) Fe$_{1-x}$Ni$_x$OOH co-catalyst overlayer,[27,28] the onset potential can be shifted to 0.90 V vs. RHE while the plateau photocurrent is similar to the bare specimens, as shown in **Figure 3**a. The application of the co-catalyst overlayer does not result in detrimental absorption or additional optical losses, as confirmed by absorptance measurements (**Figure S6**).

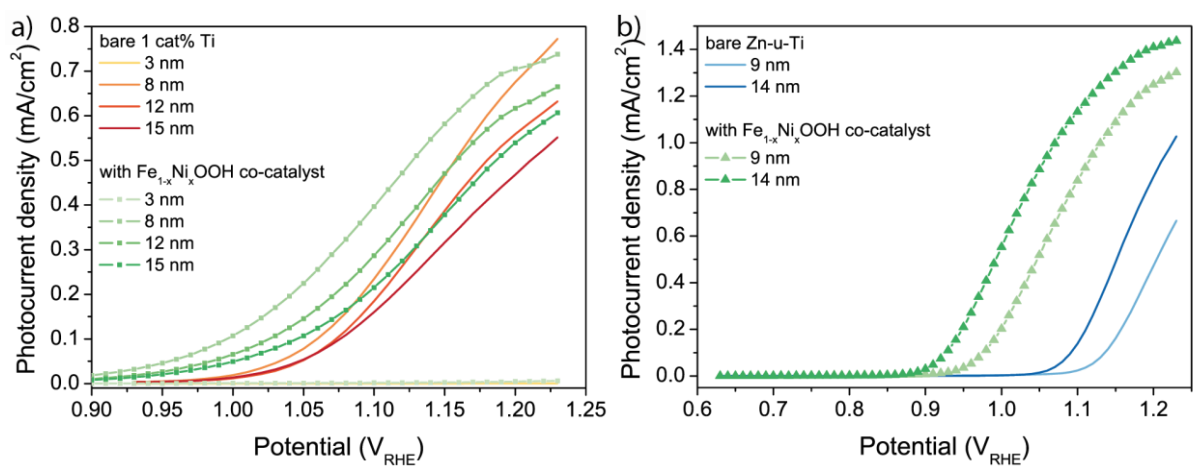

**Figure 3. Photocurrent voltammograms.** (a) Photocurrent voltammograms of homogenously 1 cat% Ti-doped hematite photoanodes without (bare hematite, orange curves) and with Fe$_{1-}$

$_x$Ni$_x$OOH co-catalyst (green curves with symbols). (b) Photocurrent voltammograms of heterogeneously doped hematite photoanodes without (bare hematite, blue curves) and with Fe$_{1-x}$Ni$_x$OOH co-catalyst (green curves with symbols).

The photocurrent decreases linearly with increasing hematite film thickness above 8 nm, see **Figure S11**b. This is somewhat surprising as the highest absorbed current density was calculated for hematite thicknesses of 20 nm (**Figure S11a**). The origin of this discrepancy most likely arises from the inability to extract the photogenerated charge carriers due to recombination in thicker films,[29] or the lack of asymmetry for charge transport and collection.[30] In order to reduce these electrical losses, heterogeneously doped hematite photoanodes with the following doping profile were fabricated: 1 cat% Zn – undoped - 1 cat% Ti doped.[13] The light and dark current densities of heterogeneously doped hematite photoanodes with hematite thickness of 9 and 14 nm are depicted in **Figure S13b**. The 9 nm-thick heterogeneously doped hematite photoanode displays a bit lower photocurrent of 0.62 mA/cm$^2$ as compared to its 8 nm-thick homogenously doped counterpart (see **Figure 3** and **Table 1**). This suggests that heterogeneous doping does not significantly assist charge separation in ultrathin films (< 10 nm). However, increasing the film thickness of the heterogeneously doped hematite to 14 nm led to a large increase in performance, achieving a photocurrent of 1.01 mA/cm$^2$ at the reversible potential of water oxidation (1.23 V vs RHE). Stable photoelectrochemical performance was demonstrated by monitoring the dark and light currents periodically over 100 h under solar simulated light at a potential of 1.13 V vs RHE (see **Figure S14**). The dark and light currents remained stable during the test (**Figure S14b**), except for sporadic fluctuations originated by O$_2$ bubbles that stuck to the opening in the photoelectrochemical cell where the photoanode was held as well as due to electrolyte evaporation during the test.

**Table 1. Optical and photoelectrochemical performance of the hematite photoanodes.** Calculated J$_{abs}$ (as shown in **Error! Reference source not found.**c), measured $J_{photo}$ at the reversible water oxidation potential (1.23 V vs. RHE) and calculated APCE = $J_{photo}$ / $J_{abs}$ (at the reversible water oxidation potential) for hematite photoanodes with different hematite layer thicknesses and doping profiles, without (bare hematite) and with Fe$_{1-x}$Ni$_x$OOH co-catalyst overlayer.

| Doping profile | 1 cat% Ti | | | | Zn-u-Ti | |
|---|---|---|---|---|---|---|
| Hematite layer thickness /nm | 3 | 8 | 12 | 15 | 9 | 14 |
| Calculated $J_{abs}$ in water /mA/cm$^2$ | 2.1 | 5.2 | 7.7 | 8.8 | 6.1 | 8.5 |
| Bare hematite | | | | | | |
| $J_{photo}$ at 1.23 V vs. RHE /mA/cm$^2$ | 0.0 | 0.8 | 0.6 | 0.5 | 0.6 | 1.0 |
| APCE at 1.23 V vs. RHE | 0.0% | 14.9% | 8.2% | 6.3% | 10.1% | 11.9% |
| With Fe$_{1-x}$Ni$_x$OOH co-catalyst | | | | | | |
| $J_{photo}$ at 1.23 V vs. RHE /mA/cm$^2$ | 0.0 | 0.7 | 0.7 | 0.6 | 1.3 | 1.4 |
| APCE at 1.23 V vs. RHE | 0.5% | 14.3% | 8.7% | 6.9% | 21.2% | 16.9% |

With the use of heterogeneously doped hematite, the benefits of the increased light absorption are evident, with the photocurrent almost doubling for hematite thicknesses of around 14-15 nm (see **Table 1**). Addition of the Fe$_{1-x}$Ni$_x$OOH co-catalyst overlayer to the heterogeneously doped hematite results in a much higher photocurrent of 1.44 mA/cm$^2$ at 1.23 V vs RHE for the 14 nm-thick hematite layer. To the authors knowledge, this is the highest photocurrent at the reversible potential reported for flat non-nanostructured ultrathin hematite photoanodes. The internal quantum efficiency or absorbed photon to current efficiency (APCE) nicely reflects the high performance of film flip and transferred specimens. The APCE was increased for a hematite thickness of ~ 14-15 nm from ~ 6 % for homogeneously doped hematite to almost ~ 12 % for the heterogeneously doped hematite (**Figure S15**). A further increase of the APCE to ~ 17 % can be achieved with co-catalyst overlayer (**Figure S15**). These results show that the film flip and transfer process used to design light trapping photonic structures with minimized optical losses have very promising performance.

## Conclusions

A film flip and transfer process was introduced to circumvent material incompatibility problems of ultrathin oxide absorbers on metallic back reflectors and leverage the full potential of the strong interference effect to enhance light-matter interaction in these nanophotonic structures. The film flip and transfer process enables depositing the metallic back-reflector (e.g. silver) after deposition of the oxide absorber, thereby avoiding tarnishing of the back-reflector that degrades its specular reflectance in the conventional bottom-up fabrication sequence. The new process was applied to fabricate ultrathin (< 20 nm) film hematite photoanodes with silver-gold alloy back-reflector for photoelectrochemical water splitting, demonstrating remarkable enhancement in light harvesting yield and photocurrent as compared to the conventional fabrication process. This case study shows the potential of the film flip and transfer process to enable fabrication of nanophotonic structures that leverage the strong interference effect for a wide range of applications. The film flip and transfer process does not introduce mechanical damage or stress induced cracking into the deposited film stack. Therefore, it is versatile, robust, and can be extended to numerous applications, opening up a new route to attach thin film stacks onto a wide range of substrates including flexible or temperature sensitive materials.

**Experimental**

Deposition

A 250 nm thick thermal $SiO_2$ layer was grown in a BTU furnace on 250 µm thick silicon substrates (University wafer) with a dry oxidation process at 1100 °C and followed by 200 nm aluminium oxide deposited in an ATC-2200 sputtering system (AJA international) on one side. An $Al_2O_3$ target (purity 99.99 %) was mounted to an RF magnetron sputtering gun. The power was set to 250 W and the sputter deposition was carried out at a pressure of 0.04 Pa with an Ar and $O_2$ gas flow of 50 sccm and 5 sccm, respectively. The substrate temperature was 23 °C and the deposition time 25000 s. Subsequently, an additional aluminium oxide film was deposited on the substrate at 700 °C by pulsed laser deposition (PLD). The ablation was performed in a PLD workstation (SURFACE systems+technology GmbH & Co. KG) with a KrF Excimer Laser (10000 pulses at 7 Hz) from a purchased target (Kurt J. Lesker). The distance between the target and the substrate was 7.5 cm. The deposition was performed in vacuum (~0.0007 Pa). The ablation of hematite photoanode and tin oxide underlayer thin films was performed in a PLD workstation (PVD Products) with a KrF Excimer Laser at 3 Hz from self-made targets (1 cat% Ti and 1 cat% Zn doped hematite)[13,15] and purchased targets (undoped hematite and $SnO_2$, Kurt J. Lesker). Depositions of the hematite and tin oxide were done at 300 °C with an oxygen flow of 3 sccm and 4 sccm at pressures of 3.33 Pa and 13.3 Pa, respectively. The hematite thickness was varied between 1500 and 6000 pulses and the $SnO_2$ underlayer was always 100 pulses. For the heterogeneously doped Zn-u-Ti structure 1 cat% Zn doped : undoped : 1 cat% Ti doped hematite were deposited from the respective targets with a pulse ratio of = 28.6% : 42.8% : 28.6%, respectively. The total number of pulses varied between 4500 and 6000 pulses. The distance between the targets and the substrate was 7.5 cm. The deposition parameters were chosen to optimize the hematite microstructure and photoelectrochemical performance.[13] Next, the stacks were annealed at 500° C for 2 h in air, with heating and cooling rate at 10° C/min. Finally, a silver-gold alloy was deposited by sputtering (10 at% Au in Ag, 99.99 % pure, Kurt J. Lesker; 0.4 Pa with 50 sccm Ar at 30 W for 3000 s).

## Film flip and transfer process

The samples were glued on another silicon substrate with a two-component glue (Epotek301, Epoxy technologies). After at least 24 h of hardening, the samples were masked with polyimide tape, and the first (carrier) silicon substrate and thermal silicon oxide layer were removed by reactive ion etching (790 series, RIE, Plasma-Therm) and deep reactive ion etching (Versaline RIE, Plasma-Therm), respectively. The etching of the first silicon oxide (etch rate 45 nm/min) was done in 36 sccm $CHF_4$ and 2 sccm $O_2$ gas flow at 5.33 Pa and 175 W for ~ 5 min. More gentle etching conditions (etch rate 11 nm/min, 5 sccm $CHF_3$ gas flow at 1.60 Pa and 97 W for ~ 15 min) were used for the second $SiO_2$. The etching of the silicon was done in cycles of A and B, repeated ~ 360 times (~ 0.56 µm/cycle, it depends on the amount of samples present). Part A: 40 sccm Ar, 60 sccm $C_4F_8$ and 0.5 sccm $SF_6$ gas flow at 2.93 Pa and 800 W for 5 min. Part B: 40 sccm Ar, 0.5 sccm $C_4F_8$ and 100 sccm $SF_6$ gas flow at 2.93 Pa and 800 W for 8 min. The aluminium oxide was removed with NaOH (1 M) in deionized water. No shear forces are present during the film flip and transfer process, which should also allow the transfer of film stacks with non-optimal adhesion. However, the top layer needs to be resistant to the used wet etchant, here NaOH solution.

For the catalyst solution $Ni(NO_3)_2 * 6H_2O$ (290.81 g/mol, 0.03 M) and $Fe(NO_3)_3 * 9H_2O$ (404.00 g/mol, 0.03 M) were dissolved in deionized water (18 Ω). The specimens were dipped for 2 min in the catalyst solution then rinsed in pure deionized water, before dipping again for 30 s in the catalyst solution and followed by a final rinse in pure deionized water.

## Characterization

The thin film surface microstructure was characterised by high resolution SEM (Zeiss Ultra-Plus FEG-SEM) with 4 kV accelerating voltages. Generally, the films were imaged by an in-lens detector and were not coated with any conductive layer.

For a selected specimen, the film microstructure was examined and the thickness was measured by cross-section TEM. The TEM cross-section lamella was prepared by the focused ion beam (FIB) etching technique (FEI Strata 400S) with a gallium liquid metal ion source, a gas injection system and a micromanipulator (Omniprobe 200). After carbon deposition by sputtering, the thin film was protected by an electron and ion beam deposition of platinum. The TEM lamella was cut free with trenches from both sides. The lamella was polished to ion transparency with currents down to 28 pA at 30 kV. Amorphization was diminished by low kV showering for several seconds at 2 kV. The thin film stack was analysed by TEM using a monochromated and aberration (image) corrected TEM (FEI Titan 80–300 kV S/TEM).

The reflectance (R) spectra of the specimens were measured using an Agilent Cary 5000 (Agilent technologies) UV/VIS spectrometer from 1200 to 300 nm. A universal measurement accessory (UMA) was used at an angle of 6° and 12° for the sample and detector, respectively.

The optical parameters (complex refractive indexes) of the layers in the final stack were obtained by spectroscopic ellipsometry (VASE Ellipsometer, J.A. Woollam). Using an in-house developed algorithm with the individual optical parameters for the different layers, we simulated the optical performance of all specimens and estimated the layer thicknesses and productive absorbance in the hematite layer. Since the reflectance was measured in air, the medium air was used for the estimation of the layer thickness while water was used as medium for the calculation of the productive absorbance and $J_{abs}$. The reported thicknesses were verified by TEM analysis. The analysis showed the thickness variation across the 2 cm$^2$ sample size is ± 1 nm.

Photoelectrochemical measurements were carried out in "Cappuccino cells", photoelectrochemical test cells equipped with a potentiostat (CompactStat, Ivium Technologies) and a solar simulator (Sun 3000 class AAA solar simulator, ABET Technologies, AM 1.5G) as a light source.[31,32] All measurements were carried out in NaOH solution (1 M, pH 13.6) made of NaOH (pearls AR, Bio-Lab ltd) and deionized water (18 Ω) and under chopped light with a short duty cycle (4 s on/6 s off). The current was measured as a function of the electrode potential using a 3-electrode setup with an Hg/HgO reference electrode and a platinum wire counter electrode. The applied potential was converted to the RHE scale using the Nernst equation. All specimens were fabricated twice and measured several times. Specimens showed repeatable performance when a maximal potential of 1.23 V vs RHE was applied. The heterogeneously doped hematite photoanodes with hematite thickness of 14 nm was tested over 100 h (see **Figure S14**). The stability test was carried out under chopped light with a short duty cycle (3 s/3 s). It still showed a current density of 0.85 mA/cm$^2$ at 1.12 V vs RHE after 100 h. The fluctuations during the measurement are due to O$_2$ bubbles formation and removing as well as evaporation and change in solvent level and temperature.

**Acknowledgements**

The research leading to these results has received funding from the European Research Council under the European Union's Seventh Framework Programme (FP/2007–2013)/ERC Grant Agreement no. 617516. B. Scherrer and D. A. Grave acknowledge support by Marie-Sklodowska-Curie Individual Fellowships no. 656132 and no. 659491, respectively. K. D. Malviya acknowledges Technion support by a fellowship from the Lady Davis Foundation. Some of the experiments reported in this work were carried out using central facilities at the Technion's Photovoltaic Laboratory supported by the Nancy & Stephen Grand Technion Energy Program (GTEP) and the Russell Berrie Nanotechnology Institute (RBNI), the Micro and Nano Fabrication Unit (MNFU), and the Hydrogen Technologies Research Laboratory (HTRL) supported by the Adelis Foundation and the Solar Fuels I-CORE program of the Planning and Budgeting Committee and the Israel Science Foundation (Grant No. 152/11). The authors would like to thank Dr. Alexander Mehlman for the support with the PLD system and Dr. Guy Ankonina for helping with the optical measurements and the aluminium oxide deposition by sputtering. The authors acknowledge technical support by the Technion's Electron Microscopy Center (MIKA).

**Contributions**

B.S. and A.K performed the research, fabricated the samples and analysed them electrochemically, optically and by SEM. B.S. wrote the manuscript. K.M. performed the TEM analysis, Y.P. the optical analysis and D.G. assisted with the co-catalyst deposition. A.R. provided the research environment. H.D. and A.R. provided conceptual advice. All authors discussed the results and their interpretation as well as critically revised the manuscript.

**Competing financial interests**

The authors declare no competing financial interests.